\documentclass[reprint,twocolumn,showpacs,superscriptaddress,groupedaddress]{revtex4-1}

\usepackage{multirow}
\usepackage{mathrsfs,amsmath}
\usepackage{verbatim}
\usepackage{amsmath}
\usepackage{amsfonts}
\usepackage{amssymb}
\usepackage{amsthm}
\usepackage{bm}
\usepackage{xparse}
\usepackage{graphicx}
\usepackage{xcolor}
\usepackage{textcase}
\usepackage{url}
\usepackage{lipsum}
\usepackage{appendix}
\usepackage{epstopdf}
\usepackage{enumitem}
\usepackage{dsfont}
\newcommand{\ket}[1]{| {#1} \rangle} % for Dirac bras
\DeclareDocumentCommand{\Tr}{m m O{\big}}{{\rm Tr}_{\:\!{#1}}#3({#2}#3)}

\begin{document}
\title{Comment on Healey's ``Quantum Theory and the Limits of Objectivity''}
\author{Veronika Baumann}
\altaffiliation[]{These authors contributed equally to this work.}
\affiliation{Faculty of Informatics, Universit\`a della Svizzera italiana, Via G. Buffi 13, CH-6900 Lugano, Switzerland}
\affiliation{Vienna Center for Quantum Science and Technology (VCQ),Faculty of Physics, University of Vienna, Boltzmanngasse 5, A-1090 Vienna, Austria}
\author{Flavio Del Santo}
\altaffiliation[]{These authors contributed equally to this work.}
\affiliation{Vienna Center for Quantum Science and Technology (VCQ),Faculty of Physics, University of Vienna, Boltzmanngasse 5, A-1090 Vienna, Austria}
\affiliation{Institute of Quantum Optics and Quantum Information (IQOQI), Austrian Academy of Sciences, Boltzmanngasse 3, A-1090 Vienna, Austria}
\author{\v{C}aslav Brukner}
\affiliation{Vienna Center for Quantum Science and Technology (VCQ),Faculty of Physics, University of Vienna, Boltzmanngasse 5, A-1090 Vienna, Austria}
\affiliation{Institute of Quantum Optics and Quantum Information (IQOQI), Austrian Academy of Sciences, Boltzmanngasse 3, A-1090 Vienna, Austria}
\date{\today}

\begin{abstract}
In this comment we critically review an argument against the existence of objective physical outcomes, recently proposed by R. Healey \cite{healey}. We show that his gedankenexperiment, based on a combination of ``Wigner's friend'' scenarios and Bell's inequalities, suffers from the main criticism, that the computed correlation functions entering the Bell's inequality are in principle experimentally inaccessible, and hence the author's claim is not verifiable. We discuss perspectives for fixing that by adapting the proposed protocol and show that this, however, makes Healey's argument virtually equivalent to other previous, similar proposals that he explicitly criticises.
\end{abstract}

\maketitle

\subsection*{Introduction}

In a recent paper \cite{healey}, R. Healey proposes an argument to supposedly show that there exist situations in which quantum measurements cannot have ``a unique, objective, physical outcome''. At the same time the author critically analyses two similar arguments put forward in \cite{brukner} and \cite{renner}, concluding that their ``dependence on questionable implicit assumptions severely limits their significance''.
Healey's argument, however, contains a series of problems and attempting to resolve them leaves it prone to the same type of criticism raised about \cite{brukner}.

The argument is based on the computation of a set of correlation functions that violate a Bell-like inequality. This violation should signify the non-existence of objective physical outcomes in quantum mechanics. The argument involves a sequence of successive measurements that have been performed by different observers. Two of these measurements are performed by observers Carol and Dan (who take the role of "Wigner's friends" \cite{wigner}), the others by "superobservers" Alice and Bob, who describe the ``friends' ''measurements unitarily and are capable of undoing them.

%{\color{red}It ought to be stressed that since the argument consists of a modification of a Bell's inequality (of CHSH-type) in a Wigner's friend scenario, it tacitly rests upon the standard assumptions of ``universal validity of quantum theory'' \cite{footnote1}, ``locality'' and ``freedom of choice''. Nonetheless, Healey has explicitly claimed that his conclusion ``is not derived using any locality assumption'' \cite{ijqf}. We deem this untenable since the argument clearly shares the same fundamental assumptions as any Bell-type argument.} However, we are primarily concerned with the methodological legitimacy of Healey's argument.\\

In deriving his no-go theorem, Healey uses the standard quantum formalism to compute expressions for correlation functions, but gives no prescription on how to relate these expressions to observable quantities. Moreover, if one assumes the standard prescriptions of quantum theory to relate all correlations entering the Bell-like inequality with recorded counts, these counts would not give rise to violation of the inequality in the proposed protocol. Hence, in the protocol it is not possible, even in principle, to verify Healey's claim against the existence of objective physical outcomes in quantum mechanics. In this respect, we would like to stress that a physical theory is composed of theoretical symbols provided with a consistent calculus to combine them (the \emph{formalism}), together with a series of \emph{rules of correspondence}. In the words of Jammer \cite{jammer}, \emph{``The formalism F, the logical skeleton of the theory, is a deductive, usually axiomatised calculus devoid of any empirical meaning; [...] To transform F into a hypothetic deductive system of empirical statements and to make it thus physically meaningful, some of the nonlogical terms, or some formulae in which they occur, have to be correlated with observable phenomena or empirical operations. [...] F without R is a meaningless game with symbols''}.\\

The protocol proposed in Ref. \cite{healey} faces three main problems:\\

(i) %No observer, in particular no subset of observers that are involved in the argument, can test (not even in principle) the violation of the proposed Bell's inequality as no one has access to full experimental data from which the correlation functions can be extracted. This applies in particular to the "superobservers" Alice and Bob who have no access to the counts registered in Carol's and Dan's measurements.\\
The violation of the proposed Bell's inequality cannot, not even in principle, be tested, because in no region of space-time are the experimental data from which all the correlation functions can be extracted available. Hence, no observer can ever test the inequality. This applies in particular to the "superobservers" Alice and Bob who have no access to the counts registered in Carol's and Dan's measurements.\\

(ii) If one wants to make the argument testable, one needs to attribute an operational meaning to the computed expressions for the correlation functions. In standard quantum mechanics, the correlation functions correspond to the relative number of counts in respective measurements. However, in the protocol of Ref. \cite{healey} Alice and Bob could only evaluate all correlation functions, if they would register the counts from Carol's and Dan's measurements (i.e. if Alice and Bob actually perform measurements). In that case they need to use the standard ``state-update rule'' for the prediction of their subsequent measurements in contrast to the assumption of Ref. \cite{healey}, wherein a measurement is assumed to proceed in accordance with a unitary interaction \cite{footnote2}. However then, there would be no violation of the Bell's inequality. %Moreover, this would contradict Healey's original assumption ``that a measurement causes no physical `collapse' of the quantum state, so that each spin-measurement proceeds in accordance with a unitary interaction between the measured particle and the rest of the experimenter's lab, and that this is consistent with its having a definite, physical outcome recorded by the experimenter in that lab'' \cite{healey}.
\\

(iii) If Alice and Bob were to verify their predictions and actually violate a Bell's inequality with collected data, one would have to adapt the protocol in a way that renders the setup analogous to that proposed in Ref. \cite{brukner}. In that case, however, one further has to require the standard assumptions for testing the violation of Bell's inequality, namely ``freedom of choice'' and ``locality''.
%in order to sensibly define a correlation function between Alice and Bob, $E(a,b)$, in the protocol in \cite{healey} one has to use assumption A1.\\
%In what follow we revise Healey's Gedankenexperiment and we discuss three main criticalities.

\subsection*{Description of the Protocol}
Similarly to proposals \cite{brukner} and \cite{renner}, the \textit{gedankenexperiment} introduced in \cite{healey} involves two ``Wigners'' (Alice and Bob) and their respective ``friends'' (Carol and Dan). The latter each receive one spin-1/2 particle (whose states are labeled by 1 and 2, respectively) of a maximally entangled pair. The protocol describes a series of operations applied by the four experimenters from the point of view of the two superobservers.
 \\

With respect to Alice's lab, the protocol is defined by the following steps:
\begin{description}[align=left]
\item[0] State preparation:\\ $|\psi (0) \rangle =  \frac{1}{\sqrt2}\left( |\uparrow \rangle_1 | \downarrow \rangle_2 - | \downarrow \rangle_1 | \uparrow \rangle_2 \right) | r \rangle_C | r \rangle_D$. Where $| r \rangle_C$ and $| r \rangle_D$ are the initial (``ready'') states of Carol and Dan, respectively, that will record the outcome of their spin measurements (and therefore do not need to be initially further specified).
\item[1]  Dan measures: $ |\psi (0) \rangle \rightarrow |\psi (1) \rangle =  U_D | \psi (0) \rangle$; where $U_D |\downarrow_d \rangle_2 | r \rangle_D =  | \text{``D sees 2} \downarrow_d \text{''}\rangle_{2D}$ (and equivalently for $ |\uparrow_d \rangle_2$).
\item[2] Carol measures: $ |\psi (1) \rangle \rightarrow |\psi (2) \rangle =  U_C | \psi (1) \rangle$; where $U_C |\downarrow_c \rangle_1 | r \rangle_C =  | \text{``C sees 1} \downarrow_c \text{''}\rangle_{1C}$ (and equivalently for $ |\uparrow_c \rangle_1$).
\item[3] Alice undoes Carol's measurement by applying $U_C^{\dagger}$: \\ $ |\psi (2) \rangle \rightarrow |\psi (3) \rangle=$\\$=  \frac{1}{\sqrt2}\left( |\uparrow_d \rangle_1 | ``\text{Dan sees 2} \downarrow_d\text{''} \rangle_{2D}\right.$\\$\left.- | \downarrow_d \rangle_1 | ``\text{Dan sees 2} \uparrow_d\text{"} \rangle_{2D} \right) | r \rangle_C =\ket{\psi(1)}$.
\item[4] Alice measures, corresponding to a unitary $U_A$. %(instead of a projection operator $|a \rangle \langle a|_1$). \\
Analogous to $U_C$, the unitary $U_A |\downarrow_a \rangle_1 | r \rangle_A =  | \text{``A sees 1} \downarrow_a\text{''}\rangle_{1A}$ (and equivalently for $ |\uparrow_a \rangle_1$)
\item[5] Bob undoes Dan's measurement by applying $U_D^{\dagger}$.
\item[6] Bob measures corresponding to a unitary $U_B$, %(instead of a projection operator $\proj{b}_2$)
which is defined as
$U_B |\downarrow_d \rangle_2 | r \rangle_B =  | \text{``B sees 2} \downarrow_b\text{''}\rangle_{2B}$ (and equivalently for $ |\uparrow_b \rangle_2$) resulting in the state $|\psi(6)\rangle = U_A U_B |\psi (0) \rangle $.
\end{description}
%Moreover, Healey notices that if Alice's lab (including Carol) is in relative motion with constant speed with respect to Bob's lab (including Dan), there is a reference frame in which ``Carol's measurement precedes Dan's, and Bob's precedes Alice's''.  Thus, with respect to Bob's lab, the protocol reads:
Healey claims that this protocol allows to predict, in every run, the correlation function between the outcomes of different observers.\\
After step $\mathbf{2}$,  Alice predicts 
\begin{equation}
E(c,d)=-\cos(c-d) \; ,
\label{coCD}
\end{equation}
where $c-d$ is the relative angle between measurement directions $\vec{c}$ and $\vec{d}$. \\
After step $\mathbf{4}$, she predicts
\begin{equation}
E(a,d)=-\cos(a-d)\; .
\label{coAD}
\end{equation}

With respect to Bob's laboratory, in relative motion with respect to Alice's (see section ``Analysis of the protocol'' for a discussion), the protocol looks as follows. 
\begin{description}[align=left]
\item [0*] State preparation:\\ $|\psi (0) \rangle =  \frac{1}{\sqrt2}\left( |\uparrow \rangle_1 | \downarrow \rangle_2 - | \downarrow \rangle_1 | \uparrow \rangle_2 \right) | r \rangle_C | r \rangle_D$.
\item [1*]  Carol  measures: $  |\psi (0) \rangle \rightarrow |\psi(1^*) \rangle = U_C | \psi (0) \rangle$.
\item [2*] Dan measures: $ |\psi(1^*) \rangle \rightarrow |\psi(2^*) \rangle =  U_D | \psi (1^*) \rangle$.
\item [3*] Bob undoes Dan's measurement by applying $U_D^{\dagger}$: \\ $ |\psi (2^*) \rangle \rightarrow |\psi (3^*) \rangle =  \ket{\psi(1^*)} $.
\item [4*] Bob measures, which is described by a unitary $U_B$.
\item [5*] Alice undoes Carol's measurement by applying $U_C^{\dagger}$.
\item [6*] Alice measures corresponding to a unitary $U_A$, resulting in the final state  $|\psi(6^*)\rangle = U_A U_B |\psi (0) \rangle $.
\end{description}
After step $\mathbf{4^*}$, Bob predicts
\begin{equation}
E(c,b)=-\cos(b-c)\; .
\label{coBC}
\end{equation}
Finally, after step $\mathbf{6}$, Alice (or alternatively Bob, after step $\mathbf{6^*}$) can compute the correlation
\begin{equation}
E(a,b)=-cos(a-b)
\label{coBA}
\end{equation}
for their observed outcomes $a$ and $b$.\\

The main feature claimed by Healey \textemdash which supposedly improves the scheme in Ref.  \cite{brukner}\textemdash \ is that in the setup of \cite{healey} there exist, for every run of the protocol, four regions of space-time where each of the observables $A_a$, $B_b$, $C_c$ and $D_d$ takes a unique, definite value (although admittedly they never co-exist, in so far as the measurement outcomes of the observers Carol and Dan are erased by the superobservers Alice and Bob). These regions are illustrated in Fig. \ref{fig1}. 
For the sake of clarity and convenience for our proposed modification of the protocol ---and in accordance with Healey's statement that the two laboratories are moving apart with constant velocities --- we represent the space-time diagram in Fig. \ref{fig1} with slight differences from the original in Ref. \cite{healey}. This leads to swapping points {\bf 1} (respectively {\bf 1*}) with  {\bf 2} (respectively {\bf 2*}) interchanging the order of Carol's and Dan's measurements in both reference-frames, but does not affect tho argument in any way.

\subsection*{Analysis of the protocol}
We present here what we consider the fundamental problems with the proposal in \cite {healey} and discuss perspectives for fixing them.\\

\textbf{1. Unitary measurements and evaluating probabilities.} 
%Healey's proposed thought experiment assumes that all measurements are unitary evolutions while relying on predictions \eqref{coCD}-\eqref{corr}, which are derived from the standard quantum formalism comprising the state-update rule.
%Arguments for using the Born- and state-update rules within fully unitary quantum theory have been made elsewhere \cite{saunders}, but they refer to agents performing measurements and collecting statistics. This means that Alice and Bob could model \emph{their} respective measurements with projectors and confirm expression \eqref{corr}. The other three correlation functions, however, are in principle 
In order to test whether quantum mechanics violates the objectivity of physical outcomes in the sense of Healey, one would need to find an operational way that would enable, at least in principle, to collect statistics from which the correlation functions could be computed and inserted into the Bell's expression.

In the protocol there is no space-time region where the measurement counts for all four measurement settings, which would enable the computation of the four correlations functions, are accessible in principle. Specifically, $E(a,d)$ is only  accessible to Alice and Dan, $E(c,b)$ only to Carol and Bob, $E(c,d)$ only to Carol and Dan, and finally $E(a,b)$ only to Alice and Bob. In three of the four calculated correlation functions either Alice or Bob or both of them perform no measurement and hence obtain no data. 
Moreover, the three correlation functions remain to be inaccessible in principle to Alice and Bob after they undo their ``friends'~'' measurements and, hence, there will be no records of Carols's and Dan's results and no means for Alice and Bob to ever evaluate any of \eqref{coCD}, \eqref{coAD} and \eqref{coBC}. 

\begin{figure}[h!]
\centering
\includegraphics[width=8.5cm]{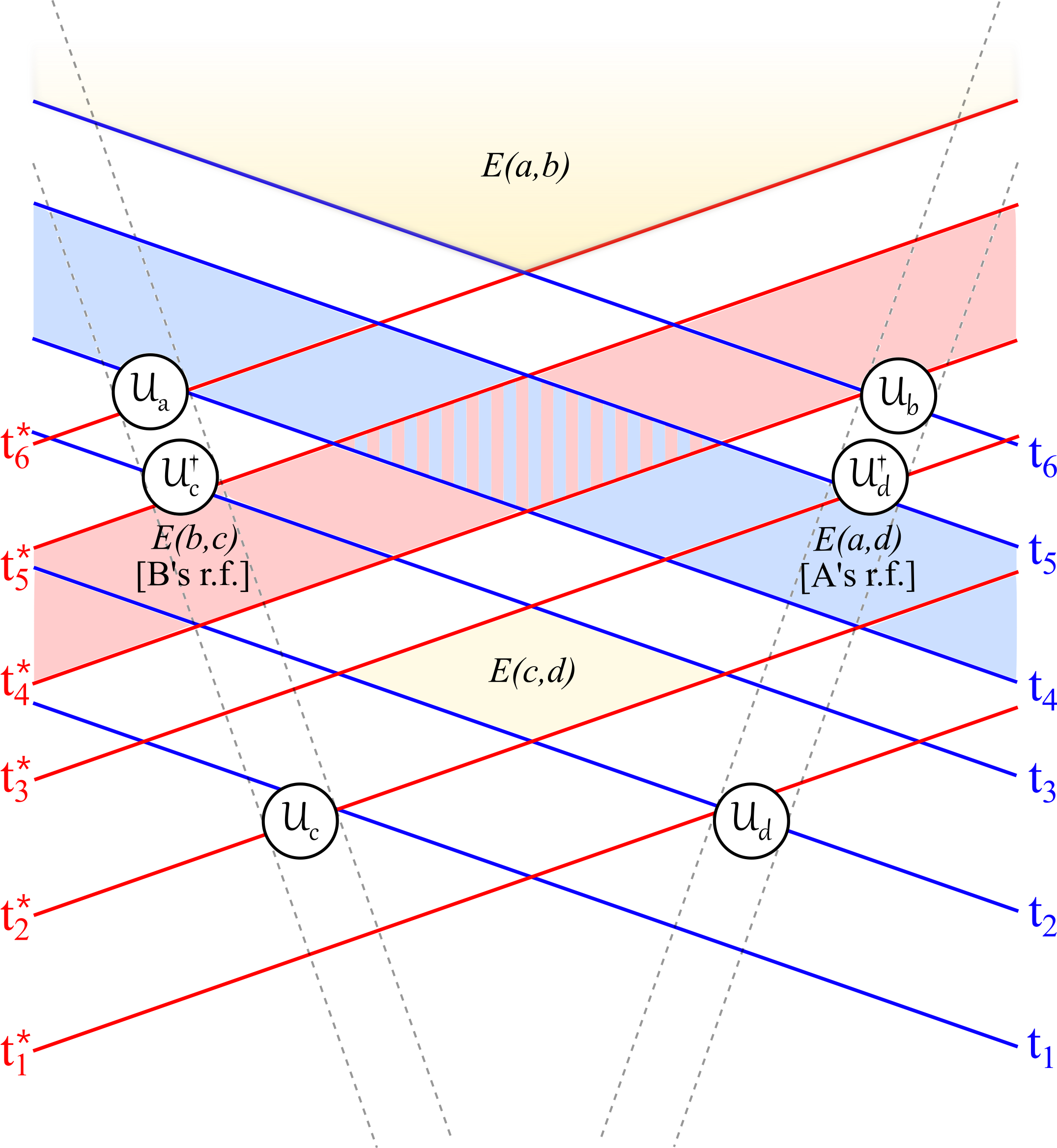}
\caption{\textbf{Space-time diagram of Healey's protocol.} Two laboratories (containing Alice-Carol-system 1 and Bob-Dan-system 2, respectively) are moving apart with constant velocity. Four agents perform a series of measurements (the order of $U_c$ and $U_d$ has been changed from the original protocol to avoid misunderstanding and to conform to the modified protocol of Fig. \ref{fig2}). In every run of the protocol there supposedly exist four space-time regions wherein correlations between the results could be in principle established. The areas highlighted in yellow are those where correlations can be established in both reference frames, whereas the blue (red) areas are those where correlations are relative to Alice's (Bob's) reference frame only.}
\label{fig1}
\end{figure}

%But even if one would assign some meaning (obviously not an operational one) to the computed correlation function questions arise.
The lack of the link between computed correlation functions and observable quantities give rise to further inconsistencies in the argument. The correlation functions $E(a,d)$ and $E(c,b)$ are computed from two different reference frames, namely the one of Alice and of Bob respectively, see Fig \ref{fig1}. In fact, it is known that the notion of quantum state, and hence of entanglement as well as the correlation functions are reference-frame dependent. This is a consequence of the relativity of simultaneity in different reference frames \cite{simulta}. %Measuring a correlation function constitutes measuring two observables at time $t=const$. This joint measurement, with respect to some frame $A$, does not constitute the measurement of a correlation function in another reference frame $B$, where the two observables are measured at different times. Conversely a correlation measurement in $B$ at $t*=const.$ will in general not correspond to a correlation measurement in $A$, due to the difference in the planes of simultaneity.
It is thus not clear to us what combining expression \eqref{coCD}-\eqref{coBA}, which are computed for different reference frames and thus for different quantum states, in a CHSH-like inequality is supposed to signify.

%
%%%%%%%%%%%%%%%%%%%%%%%%%%
%%%%%%%%%%%%%%%%%%%%%%%%%%
%

Moreover, if one would accept that combining expressions from different reference frames is non-problematic, one could equally assign different values to the correlation functions by an argument similar to that proposed in Ref.~\cite{gao}. More concretely, with respect to Alice's reference frame, the calculated correlation function $E(c,b) = 0$ for all times. For times between $t_1$ and $t_3$ the register $b$ is in a fixed pre-measurement state $|r\rangle_b$ and the results $c=+1$ and $c=-1$ occur with equal probability, whereas for times between $t_3$ and $t_6$ the result of $c$ is erased and the register is reset to it initial state $|r\rangle_c$ and the results $b=+1$ and $b=-1$ occur with equal probability. The mean value of the product of $c$ and $b$ (i.e. the correlation function) is zero independently of choice of the value assigned to the fixed states of the registers.
%the result $c$ is erased and the register is reset to its initial value corresponding to states $\ket{r}_C$ 
%before $b$ is measured. Making an analogous assignment for Bob's pre-measurement state $\ket{r}_B$, the results $\pm c$ (between $t_1$ and $t_3$) and $\pm b$ (after $t_6$) respectively will coincide with the constant values representing the ''ready''-states. This gives predicted correlation function $E(c,b)=0$ with respect to her reference frame.
Similarly, Bob would predict $E(a,d)=0$, since in his reference frame result $d$ is erased before $a$ is measured. Combining either of the two with \eqref{coCD}, \eqref{coBA} and either \eqref{coAD} or \eqref{coBC} respectively still gives a violation of the CHSH-inequality, but with a different value of $3/\sqrt{2}$. However, combining the predictions of both reference frames, as done in \cite{healey}, but using both $E(c,b)=0$ and $E(a,d)=0$ instead of \eqref{coAD} and \eqref{coBC} results in no violation of Bell's inequality:
\begin{equation}
\begin{aligned}
|E(a,b)+E(b,c)+E(c,d)-E(a,d)|=\\
|0+E(b,c)+0-E(a,d)|\leq2.
\end{aligned}
\label{CHSH}
\end{equation}
Without specifying how the computed expressions are related to observable quantities, there seems to be no physical motivation to favor the one combination of values which leads to the alleged violation, over the one that predicts no violation of the inequality.  Clearly, all these inconsistencies appear because the computed expressions for correlation functions are not associated to observable quantities. If the four measurements would be identified with four space-time points in which counts are registered, then the correlations between these counts will be reference frame independent and there would be no inconsistencies. However, then no violation would be observed in the protocol.

\textbf{2. A modified protocol that gives the expected probabilities suffers the same criticism that Healey raises.}
%In order to unambiguously recover the correlations in \eqref{corr}, one would have to adapt the protocol such that the undoing of the friend's measurements is performed by both Alice and Bob before their projective measurements in both reference frames (i.e. swapping steps 4 and 5 and $4^*$ and $5^*$, in the respective protocol). In that case, however, it becomes in principle impossible to measure any of $E(c,d)$, $E(a,d)$ and $E(b,c)$ in the same run of the protocol.
The protocol proposed in \cite{healey} allows to observe in principle only the correlations between $a$ and $b$ (since at the end of the protocol the measurement results of Carol and Dan have been erased). For the "superobservers" Alice and Bob to evaluate all the correlation functions and test the proposed CHSH-like inequality in \cite{healey}, one is forced to adapt the protocol to one wherein Alice can choose to measure either $a$ or $c$ and Bob either $b$ or $d$ by deciding whether or not to erase the measurement outcome of their ``friends''. In the latter, each ``superobserver'' (Alice and Bob) performs the same measurement as their associated observer (Carol and Dan, respectively). Since in that case one allows for a ``choice of setting'' one has to ensure ``locality'' and ``freedom of choice'' as required in a standard Bell-inequality setup. In the original protocol of Healey, ``locality'' is not required, since all four  measurements are fixed and performed in each round. Any assignment of four definite outcomes of course satisfies a Bell-like inequality by construction.\\

\begin{figure*}[h]
\centering
\includegraphics[width=17cm]{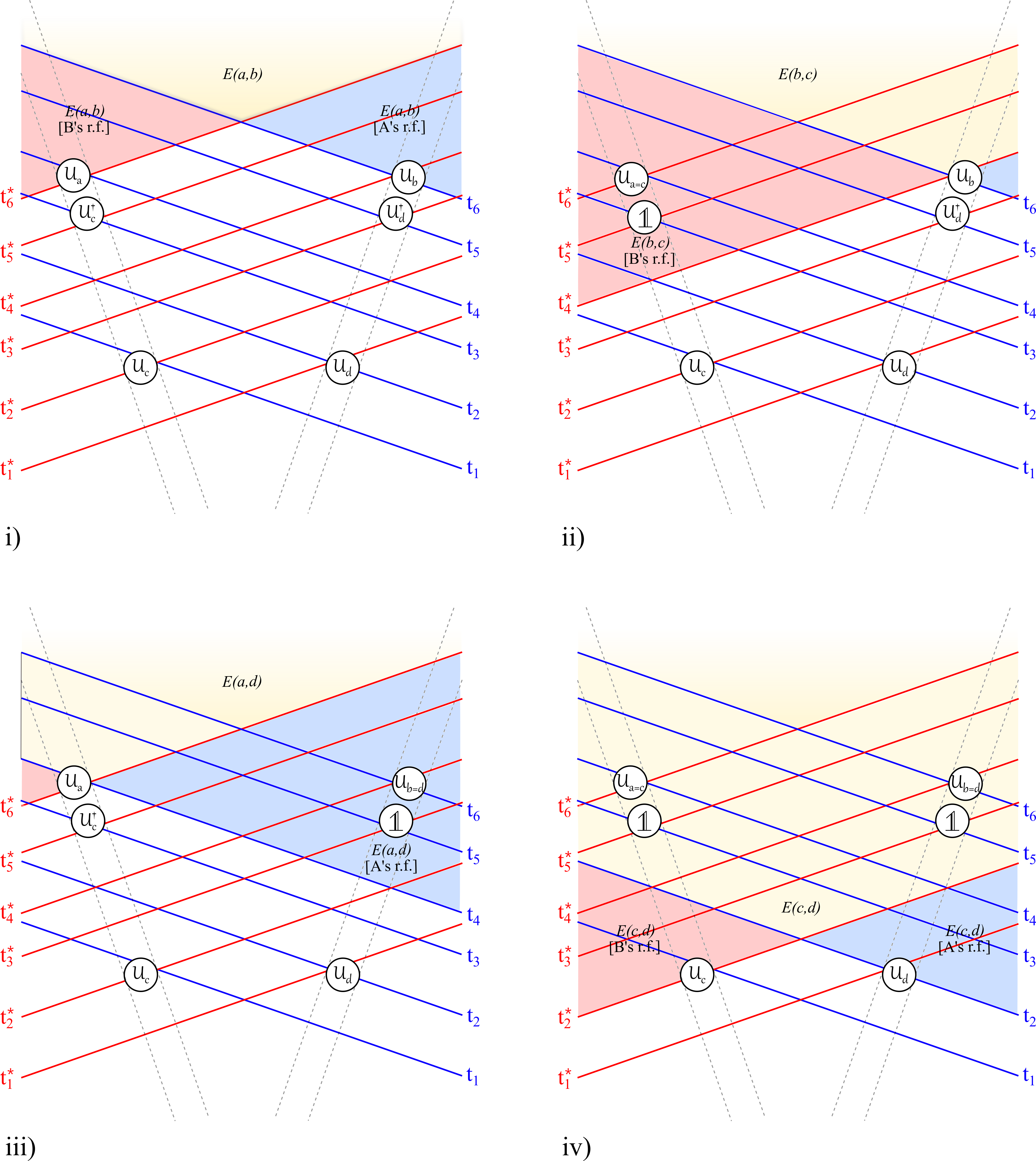}
\caption{\textbf{Space-time diagrams of the modified protocol.} The modified protocol allows the observers to chose which operations to perform on their local system in each run. This allows to provide an operational meaning to the correlations between measured quantities. At the end of the protocol the highest-level observers, Alice and Bob, unambiguously agree on the correlation functions: i) $E(a,b)$, ii) $E(b,c)$, iii) $E(a,d)$, iv) $E(c,d)$. The colour legend is defined in Fig. \ref{fig1}.}
\label{fig2}
\end{figure*}

%However, the protocol defined by the steps 0-6 allows to compute in principle only the correlations between $A_a$ and $B_b$ (since at the end of the protocol the measurement results of Carol and Dan have been erased). Therefore, in order to compute the correlations to test the proposed CHSH-like inequality, the protocol should be adapted to one wherein Alice and Bob can decide whether not to erase the measurement outcome of their ``friends'', by measuring in the same directions of Carol and Dan, respectively
In the adapted protocol, Alice \emph{not} erasing Carol's measurement, but rather measuring Carol's observable, allows for computing $E(b,c)$, whereas Bob \emph{not} erasing Dan's measurement, but performing the same measurement instead will make $E(a,d)$ experimentally accessible; whereas, both the ``superobservers'' \emph{not} undoing their friends' measurements allows for measuring $E(c,d)$. This adapted protocol is illustrated in Fig. \ref{fig2}.
Note that in the rounds where Alice, Bob or both decide to measure the observables of Carol or Dan or both respectively, and register counts in the respective measurements, they cannot erase these counts in a unitary fashion to continue with the subsequent measurements. %This is the point in the protocol that displays the fundamental problem with assumption $\mathcal A$. 
Upon registering the counts in the first measurements, Alice and Bob will observe a statistics in the subsequent measurement, which is compatible with the state-update rule and not with a unitary transformation. The continuation of the protocol in its original form \cite{healey} but with the application of the state-update rule would not lead to a violation of Bell's inequalities. It appears therefore that the only way to give an operational meaning to accessing the correlation functions and hence to the verification of Healey's argument is to assume that in each run Alice and Bob chooses one of the two measurement settings \textemdash a step that was criticised by Healey.
\\
%%%%%%%%%%%%%%%%%%%%%%%%%%
%%%%%%%%%%%%%%%%%%%%%%%%%%
%Note that in the rounds where Alice, Bob or both decide to not to erase the friend's measurements, the outcomes $a$ and $b$ do not co-exist accordingly.

It, therefore, seems to us that this setup is subject to the same criticism as \cite{brukner} when one requires the actual computation of the terms violating the CHSH-inequality to correspond to measured values. Otherwise, we think equations \eqref{coCD}-\eqref{coBA} need further justification.\\

In conclusion, the only possibility for the correlations to be in principle measured is to revise the scheme in a way that does not allow to be sure that the four values coexisted in each round, therefore suffering the very same criticism that was addressed towards proposal \cite{brukner}. However, one can make a stronger assumption that these values are fixed even when they are not all co-measured, which then allows to derive the no-go theorem of Ref. [2].\\

\begin{acknowledgements}
We would like to thank Richard Healey and Shan Gao for useful comments and discussion. FDS would like to acknowledge the financial support through a DOC Fellowship of the Austrian Academy of Sciences. VB wants to acknowledge the financing by Swiss National Science Foundation (SNF) and the NCCR QSIT. \v{C}B acknowledges the support of the Austrian Science Fund (FWF) through the project I-2906 and the support of a grant from the John Templeton Foundation. The opinions expressed in this publication are those of the authors and do not necessarily reflect the views of the John Templeton Foundation.
\end{acknowledgements}

\begin{small}

\end{small}


\begin{thebibliography}{200}
\bibitem{footnote2} In Ref. \cite{healey}, Healey states ``that a measurement causes no physical `collapse' of the quantum state, so that each spin-measurement proceeds in accordance with a unitary interaction between the measured particle and the rest of the experimenter's lab, and that this is consistent with its having a definite, physical outcome recorded by the experimenter in that lab''
\end{thebibliography}
\end{document}